\documentclass{article}
\usepackage{amssymb}
\usepackage{amsthm}

\newtheorem{proposition}{Proposition}
\newtheorem{definition}{Definition}
\newtheorem{example}{Example}

\title{On totally umbilic submanifolds of \\ semi-Riemannian manifolds}

\author{Volker Perlick \\ {\small Institut f\"ur Theoretische Physik,
Sekr. PN 7-1} \\ {\small TU Berlin, 10623 Berlin, Germany} \\ 
{\small Email: vper0433@itp.physik.tu-berlin.de}}%
\date{}       
\begin{document}

% typeset front matter
\maketitle

\begin{abstract}
The notion of being totally umbilic is considered for non-degenerate
and degenerate submanifolds of semi-Riemanian manifolds. After some
remarks on the general case, timelike and lightlike totally umbilic 
submanifolds of Lorentzian manifolds are discussed, along with their 
physical interpretation in view of general relativity. In particular,
the mathematical notion of totally umbilic submanifolds is linked to
the notions of photon surfaces and of null strings which have been
used in the physics literature.
\end{abstract}

%%%%%%%%%%%%%%%%%%%%%%%%%%%%%%%%%%%%%%%%%%%%%%%%%%%%%%%%%%%%%%%%%%%%%%%%%%%%%%%%%%%%%%%%%%%%%

\section{Introduction}
\label{sec-intro}
In a Riemannian manifold, a submanifold is said to have an \emph{umbilic
point} at $p$ if, at this point $p$, the second fundamental form is a
multiple of the first fundamental form. A submanifold is called \emph{totally
umbilic} if all of its points are umbilic. In $n$-dimensional Euclidean 
space, a $k$-dimensional complete connected submanifold with $2 \le k \le n-1$
is totally umbilic if and only if it is a $k$-sphere or a $k$-plane.

If we generalize from Riemannian to semi-Riemannian manifolds (i.e., if
the metric of the ambient space need not be positive-definite), a submanifold 
may be degenerate (i.e., the metric may induce a degenerate tensor field on 
this submanifold). In this case the standard text-book definition of the second 
fundamental form (or \emph{shape tensor}) does not make sense; so many authors 
restrict the definition of totally umbilic submanifolds to the
non-degenerate case, see, e.g., O'Neill \cite{o}. However, there is a
fairly obvious way in which the definition of the second fundamental
form can be generalized to include degenerate submanifolds. This is
indicated in Exercise 9 on p.125 in O'Neill's book \cite{o} but not used in the
main part. The geometry of degenerate submanifolds of semi-Riemannian
manifolds is discussed in detail in the books by Kupeli \cite{k}
and by Duggal and Bejancu \cite{db}

In this article I use the generalized definition of the second fundamental
form, as suggested in the above-mentioned exercise of O'Neill's book,
and discuss, thereupon, some general properties of totally umbilic
submanifolds which may or may not be degenerate. In the second part of the
article I specialize to the case that the ambient space has Lorentzian
signature and give several characterizations of totally umbilic submanifolds
that are timelike or lightlike. The interpretation of these results in
view of general relativity is also discussed. It is one of the main purposes 
of this article to link up the mathematical literature on totally umbilic
submanifolds with some notions used in the physical literature; these are,
in particular the notions of \emph{photon surfaces} (introduced by Claudel,
Virbhadra and Ellis \cite{cve} and also discussed by Foertsch, Hasse and
Perlick \cite{fhp}) and of \emph{null strings} (introduced by Schild
\cite{sch}). 

%%%%%%%%%%%%%%%%%%%%%%%%%%%%%%%%%%%%%%%%%%%%%%%%%%%%%%%%%%%%%%%%%%%%%%%%%%%%%%%%%%%%%%%%%%%%%

\section{Definition and general properties of totally umbilic submanifolds
in semi-Riemannian manifolds}
\label{sec:general}
Let $(M,g)$ be a semi-Riemannian manifold and $\nabla$ the Levi-Civita
connection of $g$. On every (immersed) submanifold $\tilde{M}$ of $M$,
$g$ induces a second-rank tensor field $\tilde{g}$. If $\tilde{g}$
has a non-trivial kernel, the submanifold is called degenerate; otherwise
it is called non-degenerate. 

In the non-degenerate case $(\tilde{M},\tilde{g})$
is a semi-Riemannian manifold in its own right; in particular, it defines
a Levi-Civita connection on $\tilde{M}$. In the
degenerate case, the ambient space induces, in general, no distinguished 
connection on $\tilde{M}$

Moreover, in the
non-degenerate case at each point $p$ in $\tilde{M}$ the tangent space
splits orthogonally,
\begin{equation}\label{eq:ortho}
  T_p M = T_p \tilde{M} + T_p^{\perp} \tilde{M}
\end{equation}
where $T_p^{\perp} \tilde{M}$ denotes the set of all vectors in $T_pM$
that are perpendicular to $T_p\tilde{M}$ with respect to $g$. In the
degenerate case, (\ref{eq:ortho}) does not hold because $T_p \tilde{M}$
and $T_p^{\perp} \tilde{M}$ have a non-trivial intersection and do
not span the whole tangent space $T_pM$. In other
words, a vector in $T_pM$ cannot be decomposed uniquely into a component
tangent to $\tilde{M}$ and a component perpendicular to $\tilde{M}$.
This is the reason why the standard text-book definition of the second
fundamental form, which makes use of this decomposition, does not work.
However, this problem can be easily circumvented by using the quotient
space $T_pM / T_p \tilde{M}$ instead of $T_p ^{\perp}\tilde{M}$.
We denote the elements of this quotient space by square brackets,
i.e., we write
\begin{equation}\label{eq:quotient}
  [Z_p] = \{ Z_p + Y_p \, \vert \, Y_p \in T_p \tilde{M} \}
\end{equation}
for $Z_p \in T_pM$.  If $Z$ is a vector field along $\tilde{M}$, we denote
by $[Z]$ the map that assigns to each point $p \in T_p \tilde{M}$ the
equivalence class $[Z_p ]$, where $Z_p$ is the value of $Z$ at $p$. Using 
this notation, we define the second fundamental form $\Pi$ of $\tilde{M}$ 
by the equation
\begin{equation}\label{eq:defPi}
  \Pi (X,Y) = [ \nabla _X Y ]
\end{equation}
where $X$ and $Y$ are vector fields tangent to $\tilde{M}$. As $\nabla$ is
torsion free, $\nabla _X Y - \nabla _Y X = [X,Y]$, and as with $X$ and $Y$
also the Lie bracket $[X,Y]$ must be tangent to $\tilde{M}$, it is clear
that $\Pi$ is symmetric, $\Pi (X,Y) = \Pi (Y,X)$. This symmetry property
implies that $\Pi$ is tensorial with respect to both arguments, because it
is obviously tensorial with respect to the first one. This reasoning is 
quite analogous as for the standard text-book definition of the second
fundamental form; the only difference to this standard definition lies in the
fact that at each point $p \in \tilde{M}$ now $\Pi(X,Y)$ takes values
in $T_pM/T_p\tilde{M}$ rather than in $T_p^{\perp} \tilde{M}$. For
non-degenerate submanifolds, these two spaces can of course be identified
in a natural fashion.

With $\Pi$ given by (\ref{eq:defPi}), we can now define the notion of
being totally umbilic for submanifolds that may be degenerate or
non-degenerate.

\begin{definition}\label{def:umbilic}
  A submanifold $\tilde{M}$ of a semi-Riemannian manifold is called
  \emph{totally umbilic} if there is a vector field $N$ along $\tilde{M}$
  such that
  \begin{equation}\label{eq:umbilic}
    \Pi(X,Y) = [g(X,Y) N]
  \end{equation}
  for all vector fields $X$ and $Y$ tangent to $\tilde{M}$.
  A totally umbilic submanifold with $[N]=[0]$ is called
  \emph{totally geodesic}.
\end{definition}

Note that the property of being totally umbilic is invariant under 
conformal changes of $g$ whereas the property of being totally geodesic 
is not. Also note that for a totally umbilic submanifold the equivalence 
class $[N]$ is unique but the vector field $N$ is not. In the non-degenerate 
case we can make $N$ unique by requiring that it be perpendicular to $\tilde{M}$. 
In the totally geodesic case, we can of course choose $N=0$. For degenerate 
totally umbilic submanifolds that are not totally geodesic, however, there is 
no distinguished choice for the vector field $N$. 

Once $N$ has been chosen, the equation
\begin{equation}\label{eq:connection}
 \tilde{\nabla} _X Y = \nabla _X Y - g(X,Y) N
\end{equation}
defines a torsion-free connection $\tilde{\nabla}$ on $\tilde{M}$. For non-degenerate
totally geodesic submanifolds, $\tilde{\nabla}$ with the choice
$N=0$ coincides with the Levi-Civita connection of $(\tilde{M}, \tilde{g})$.

We now prove a proposition which is a simple consequence of the existence
of the connection (\ref{eq:connection}). In this proposition we use the 
following terminology. (This terminology comes from general relativity and is, 
actually, motivated only in the case that the metric of the ambient space 
has Lorentzian signature.) We call a vector field $X$ on $M$ lightlike if
$g(X,X)=0$; a curve is called lightlike if it is the integral curve
of a lightlike vector field.

\begin{proposition}\label{prop:geodesic}

\begin{itemize}
\item[\emph{(a)}]
  Let $\tilde{M}$ be a totally umbilic submanifold of $M$. Then a lightlike
  $\nabla$--geodesic that starts tangential to $\tilde{M}$ remains within
  $\tilde{M}$ $($for some parameter interval around the starting point$)$.
\item[\emph{(b)}]
  $\tilde{M}$ is totally geodesic if and only if every $\nabla$--geodesic that
  starts tangential to $\tilde{M}$ remains within $\tilde{M}$ $($for some
  parameter interval around the starting point$)$.
\end{itemize}
\end{proposition}
\begin{proof}
  From (\ref{eq:connection}) we read that for lightlike vector fields
  on $\tilde{M}$ the equation $\tilde{\nabla} _X X =0$ is equivalent 
  to $\nabla _X X =0$. In other words, the $\nabla$--geodesics with lightlike
  initial vectors tangent to $\tilde{M}$ are $\tilde{\nabla}$ geodesics and
  thus remain within $\tilde{M}$. This proves (a). In the totally geodesic case 
  we may choose $N=0$, so the same argument works for non-lightlike initial
  vectors as well. This proves (b).
\end{proof}

Part (b) is, of course, the true justification for the name 'totally geodesic'.

The following characterization is often useful.

\begin{proposition}\label{prop:tangent}

\begin{itemize}
\item[\emph{(a)}]
  $\tilde{M}$ is totally umbilic if and only if all vector fields
  $X$ and $Y$ which are tangent to $\tilde{M}$ with $g(X,Y)=0$
  satisfy $\Pi (X,Y) = [0]$.
\item[\emph{(b)}]
  $\tilde{M}$ is totally geodesic if and only if all vector fields
  $X$ and $Y$ which are tangent to $\tilde{M}$ satisfy $\Pi (X,Y) = [0]$.
\end{itemize}
\end{proposition}
\begin{proof}
  Claim (b) and the 'only if' part of claim (a) are obvious
  from Definition \ref{def:umbilic}. To prove the 'if' part of claim (a),
  we choose basis vector fields $E_1, \dots , E_l,L_1,
  \dots , L_m$ on $\tilde{M}$ such that the $E_i$ are pseudo-orthonormal
  and $g(L_{\mu},L_{\nu})=g(L_{\mu},E_i)=0$ for all $\mu,\nu = 1, \dots ,m$
  and all $i=1, \dots ,l$. For each $i$, we define a vector field $N_i$
  along $\tilde{M}$ by $\nabla _{E_i} E_i = g(E_i,E_i)N_i$. Considering
  for $X$ and $Y$ all linear combinations of $E_i$ and $E_j$ that are orthogonal to each
  other, our hypothesis implies that $[N_i]=[N_j]=:[N]$ for all $i,j=1,
  \dots , l$. With this information at hand, we consider for $X$ and $Y$
  arbitray vector fields tangent to $\tilde{M}$, i.e., linear combinations
  of all $E_i$ and  $L_{\mu}$; then our hypopthesis implies that $\Pi (X,Y)$
  is, indeed, of the form (\ref{eq:umbilic}).
\end{proof}

It follows directly from Definition \ref{def:umbilic} that a non-degenerate
one-dimensional submanifold is always totally umbilic and that a degenerate
one-dimensional submanifold is totally umbilic if and only if it is the
image of a geodesic. For this reason the notion of totally umbilic submanifolds
is non-trivial only for $1 < \mathrm{dim} ( \tilde{M} ) < \mathrm{dim} (M)$.
In an arbitrary semi-Riemannian manifold, the existence of non-trivial
totally umbilic submanifolds is not guaranteed. For the case that the
ambient space is Riemannian, existence criteria for totally umbilic
foliations in terms of curvature
conditions have been given by Walschap \cite{wa}. Generalizations to
the semi-Riemannian case have, apparently, not been worked out so far.
Here are two simple examples of semi-Riemannian manifolds that do
admit non-trivial totally umbilic submanifolds.

\begin{example}\label{ex:flat}
O'Neill \emph{\cite{o}}, p.$117$, considers the case where $(M,g)$ is pseudo-Euclidean,
i.e. $\mathbb{R} ^n$ with a constant metric of arbitrary signature.
He shows that for $n \ge 3$ a complete connected non-degenerate hypersurface
is totally umbilic if and only if it is either a hyperplane or a hyperquadric.
Every connected non-degenerate totally umbilic submanifold
$\tilde{M}$ with $2 \le \mathrm{dim} ( \tilde{M} ) \le \mathrm{dim} (M) -2$ 
is a hypersurface in some pseudo-Euclidean subspace of dimension
$\mathrm{dim} ( \tilde{M})+1$; this was proven by
Ahn, Kim, and Kim \emph{\cite{akk}} $($cf. Hong \emph{\cite{ho}} for the
Lorentzian case$)$. Thus, with the O'Neill result we know 
\emph{all} non-degenerate totally umbilic submanifolds of pseudo-Euclidean
space.
\end{example}

\begin{example}\label{ex:twisted}
Consider the case that $(M,g)$ is $($locally$)$ a \emph{twisted product}.
By definition, this means that $M$ admits coordinates $(u,v) = 
(u^1,\dots, u^m, v^1, \dots , v^{n-m})$ $($locally around any point$)$
such that the metric $g$ takes the form
\begin{equation}\label{eq:twist}
  g \, = \, h_{ij}(u) \, du^i \, du^j \, + \,
\psi(u,v) \, k_{\mu \nu} (v) \, dv^{\mu} \, dv^{\nu}
\end{equation}
with summation over $i,j$ from 1 to $m$ and over $\mu, \nu$ from
1 to $n-m$. The condition of $g$ being non-degenerate requires $\psi$
to be non-zero and $h_{ij}$ and $k_{\mu \nu}$ to be non-degenerate everywhere; 
otherwise, they are arbitrary. $($In the more special case that the ``twisting 
function'' $\psi$ is independent of $v$ one speaks of a {\em warped product}, 
cf. O'Neill \emph{\cite{o}}.$)$ It is an elementary exercise to verify that for a metric 
of the form \emph{(\ref{eq:twist})} the submanifolds $u = {\mathrm{constant}}$
are totally umbilic and the submanifolds $v = {\mathrm{constant}}$ are
totally geodesic. More generally, the following result is true. A 
semi-Riemannian manifold is $($locally$)$ a twisted product if and only if it 
$($locally$)$ admits two foliations ${\mathcal{F}}$ and ${\mathcal{G}}$ which are 
transverse and orthogonal to each other $($and thus non-degenerate$)$ with
all leaves of ${\mathcal{F}}$ totally geodesic and all leaves of
${\mathcal{G}}$ totally umbilic, see Ponge and Reckziegel
\emph{\cite{pr}}, Theorem $1$. Recall that the notion of being totally umbilic is
conformally invariant. So every metric that is $($locally$)$ conformal to a 
twisted product admits $($locally$)$ two foliations into non-degenerate totally 
umbilic submanifolds that are orthogonal to each other. This observation
implies that a semi-Riemannian manifold can be $($locally$)$ foliated into non-degenerate
totally umbilic hypersurfaces if and only if it admits $($locally$)$ coordinates
$(u,v)=(u,v^1, \dots , v^ {n-1})$ such that
\begin{equation}\label{eq:hypersurface}
  g \, = \, \Phi(u,v) \big( \pm du^2 + \psi (u,v) \, k_{\mu \nu } (v) \,
  dv^{\mu} \, dv^{\nu} \, \big) \: .
\end{equation}
Here we made use of the fact that, for a foliation into non-degenerate
hypersurfaces, the orthocomplements of the leaves are one-dimensional and,
thus. integrable. If we want to know if a given metric can be foliated into
non-degenerate totally umbilic hypersurfaces, we may thus do this by checking
whether it is isometric to \emph{(\ref{eq:hypersurface})}. 
\end{example}

%%%%%%%%%%%%%%%%%%%%%%%%%%%%%%%%%%%%%%%%%%%%%%%%%%%%%%%%%%%%%%%%%%%%%%%%%%%%%%%%%%%%%%%%%%%%%

\section{Totally umbilic submanifolds of Lorentzian manifolds}
\label{sec:Lorentzian}

From now on we assume that the ambient space $(M,g)$ has Lorentzian signature
$(+, \dots , + , -)$.  We may then interpret $(M,g)$ as a spacetime in the
sense of general relativity. (However, there is no need to restrict our discussion
to the physically interesting case $\mathrm{dim} (M) =4$.) As usual in the
Lorentzian case, we call the degenerate submanifolds \emph{lightlike}.
On a non-degenerate submanifold, the metric is either positive definite or again Lorentzian;
in the first case, the submanifold is called \emph{spacelike}, in the
second case it is called \emph{timelike}. It is our goal to discuss
totally umbilic submanifolds that are timelike or lightlike. (The spacelike
case is not very much different from the situation that the ambient space is
Riemannian.) We begin with the timelike case.

\begin{proposition}\label{prop:timelike}
  Let $\tilde{M}$ be a timelike submanifold with $2 \le \mathrm{dim}
  (\tilde{M}) \le \mathrm{dim} (M)$. Then $\tilde{M}$ is totally umbilic
  if and only if every lightlike geodesic that starts tangent to $\tilde{M}$
  remains within $\tilde{M}$ $($for some parameter interval around the starting
  point$)$.
\end{proposition}
\begin{proof}
  The 'only if' part is a special case of Proposition \ref{prop:geodesic} (a).
  To prove the 'if' part, choose a point $p \in \tilde{M}$ and two vectors
  $X_p$ and $Y_p$ in $T_p \tilde{M}$ with $g_p(X_p,Y_p)=0$. It is our goal
  to prove that $\Pi _p (X_p , Y_p) = [0]$ because then, by the tensorial
  property of $\Pi$, Proposition \ref{prop:tangent} (a) proves that 
  $\tilde{M}$ is totally umbilic. Owing to the tensorial property of $\Pi$,
  it suffices to consider the case that $g_p(X_p,X_p)=1$ and $g_p(Y_p,Y_p)=-1$.
  Then the vectors $L_p = \frac{1}{2} (Y_p + X_p)$ and $K_p = \frac{1}{2} (
  Y_p - X_p)$ are lightlike. By hypothesis, we can find lightlike vector fields 
  $L$ and $K$ on $\tilde{M}$ with $\nabla _L L = \nabla _K K = 0$ which take the 
  values $L_p$ and $K_p$ at $p$. Writing 
  $X=L-K$ and $Y=L+K$ we find $\Pi (X,Y) = [ \nabla _{(L-K)} (L+K)]= 
  [ \nabla _L K - \nabla _K L]$ which is, indeed, equal to $[0]$ because with 
  $L$ and $K$ also the Lie bracket $[L,K]$ is tangent to $\tilde{M}$.
\end{proof}

In general relativity, lightlike geodesics are interpreted as the
wordlines of photons. Therefore, a timelike or lightlike submanifold 
$\tilde{M}$ is called a \emph{photon surface} if each lightlike geodesic
that starts tangent to $\tilde{M}$ remains within $\tilde{M}$ (for some
parameter interval). In this terminology, Proposition \ref{prop:timelike}
says that a timelike submanifold of dimension $k \ge 2$ is
totally umbilic if and only if it is a photon surface. The
notion of a photon surface was discussed by Claudel, Virbhadra, and
Ellis  \cite{cve} for the case $k= \mathrm{dim} (M) -1$ and by Foertsch,
Hasse and Perlick \cite{fhp} for the case $k =2$.

A $k$-dimensional timelike submanifold can be interpreted as the
history of a $(k-1)$-dimensional spatial manifold. Proposition
\ref{prop:timelike} says that this spatial manifold appears like
a $(k-1)$-plane to the eye of every observer in $\tilde{M}$ if and 
only if its history is a totally umbilic submanifold. In particular,
a 2-dimensional timelike submanifold can be interpreted as
the history of a string; the condition of being totally umbilic
means that the string looks like a straight line to an observer on 
the string. The best known non-trivial example is the surface $r=3m, 
\theta = \pi /2$ in Schwarzschild spacetime; it is the history of
a circle which appears like a straight line to the eye of an
observer who is situated on this circle. Other examples are worked out
in \cite{fhp}. There the reader can also find a characterization
of 2-dimensional timelike photon surfaces in terms of inertial forces and in terms
of gyroscope transport. For axisymmetric and static situations, 
this connection was discussed already earlier in various articles 
by Abramowicz, see, e.g., \cite{a}.

If a Lorentzian manifold admits a timelike conformal Killing field $K$
that is hypersurface-orthogonal, applying the flow of $K$ to a lightlike
geodesic always gives a 2-dimensional timelike photon surface. The proof
is worked out in \cite{fhp}. This demonstrates the existence of 2-dimensional
timelike totally umbilic submanifolds in any conformally static spacetime.
Actually, for the construction to work it is not necessary that $K$ be
timelike; it suffices if it is nowhere orthogonal to the lightlike geodesic
to which we want to apply the flow of $K$. Most known examples of 2-dimensional
timelike photon surfaces are constructed in this way. However, 
this construction is not universal; with the help of Example
\ref{ex:twisted} one can construct 3-dimensional Lorentzian manifolds which
do not admit any non-zero comformal Killing vector field but are foliated
into 2-dimensional timelike photon surfaces.

We emphasize again that, in an $n$-dimensional Lorentzian manifold
the existence of a $k$-dimensional totally umbilic timelike submanifold
is not guaranteed unless in the trivial cases $k=1$ and $k=n$. This
implies that, in particular, the existence of totally geodesic
timelike submanifolds of dimension $2 \le k \le n-1$ is not guaranteed.
This has physical relevance for the case $k=2$ because 2-dimensional
timelike submanifolds that are totally geodesic describe the history of 
self-gravitating strings, see Vickers \cite{v}.

We now turn to the case of a lightlike submanifold $\tilde{M}$.
We first observe that every such $\tilde{M}$ is ruled by a unique
congruence of lightlike curves which are called the \emph{generators}
of $\tilde{M}$. This leads to the following characterization of
totally umbilic lightlike submanifolds.

\begin{proposition}\label{prop:light}
 A necessary condition for a lightlike submanifold to be totally umbilic
 is that the generators are geodesics. In a three-dimensional Lorentzian 
 manifold, every two-dimensional lightlike submanifold $\tilde{M}$ is totally 
 umbilic.
\end{proposition}
\begin{proof}
  The first statement is a special case of Proposition \ref{prop:geodesic} (a).
  To prove the second statement, let $L$ be a vector field on
  $\tilde{M}$ that is tangent to the generators and $E$ any other
  vector field on $\tilde{M}$ that is linearly independent of $L$.
  Set $\nabla _E E =: g(E,E) N$. As $g(L,L)=0$, the vector field
  $\nabla _E L$ is perpendicular to $L$ and, thus, tangent to
  $\tilde{M}$. As $\nabla _L L$ is a multiple of $L$ and $g(E,L)=0$,
  the vector field $\nabla _L E$ is also perpendicular to $L$ and, thus,
  tangent to $\tilde{M}$. As a consequence, every linear combination
  $X=fE+hL$ satisfies $\Pi (X,X) = [f^2 g(E,E) N] = [g(X,X) N]$. As
  $\Pi$ is symmetric, this proves that $\tilde{M}$ is totally umbilic.
\end{proof}

For lightlike hypersurfaces, the generators are automatically geodesics.

Schild \cite{sch} has defined a \emph{null string} as a 2-dimensional lightlike
submanifold whose generators are geodesics. Whereas every 2-dimensional lightlike
submanifold can be interpreted as the history of a string whose individual points
move at the speed of light, null strings are characterized by the additional
condition that its individual points move on geodesics, i.e., freely like photons. 
In other words, we may visualize a null string as a one-parameter family of 
photons that are arranged like perls on a string. In a three-dimensional Lorentzian 
manifold, every two-dimensional lightlike submanifold is totally umbilic and it is 
a null string in the sense of Schild.

In Schild's article it is shown that null strings can be characterized by a 
variational principle. From this variational principle it is clear that
null strings exist in every Lorentzian manifold. This can also be verified
with the help of the following construction. Choose a one-dimensional spacelike 
submanifold $S$; at each point of $S$, choose a lightlike direction perpendicular 
to $S$ that depends smoothly on the foot-point; let $\tilde{M}$ be the
union of the lightlike geodesics determined by these intial directions;
in a neighborhood of $S$, this is indeed a submanifold and,
by construction, it is a null string. (Farther
away from $S$ the set $\tilde{M}$ constructed this way may form
'caustics', i.e., it may fail to be a submanifold.) 

We can try to construct $k$-dimensional totally umbilic lightlike
submanifolds in Lorentzian manifolds of arbitrary dimension by the
same procedure, now starting with a $(k-1)$-dimensional spacelike
submanifold $S$. However, even if
we are lucky enough to find a totally umbilic initial submanifold
$S$, it will not be guaranteed that the resulting lightlike 
submanifold will be totally umbilic everywhere. We end with an
example where higher-dimensional totally umbilic lightlike submanifolds
can be constructed owing to the existence of symmetries.

\begin{example}\label{ex:spheric}
Consider a $4$-dimensional Lorentzian manifold that is spherically
symmetric and static, i.e.,
\begin{equation}\label{eq:spheric}
  g \, = \, -A(r)^2 dt^2 + B(r)^2 dr^2 + r^2 ( \mathrm{sin}^2 \vartheta
  d \varphi ^2 + d \vartheta ^2 ) \, ,
\end{equation}
e.g. the Schwarzschild metric with $A(r)^2=B(r)^{-2}=1- \frac{2m}{r}$.
Let $S$ be the intersection of a hypersurface $r = \mathrm{constant}$
with a hypersurface $t= \mathrm{constant}$. This is a $2$-dimensional
spacelike submanifold. Now choose at each point of $S$ a lightlike
direction perpendicular to $S$, smoothly depending on the foot-point.
$($You have to choose between two possibilities: the ingoing radial
directions and the outgoing radial directions.$)$ Let $\tilde{M}$ be
the union of all geodesics with the chosen initial direction. This is
a submanifold near $S$. $($In the case at hand, the symmetry of the 
situation guarantees that $\tilde{M}$ is a submanifold everywhere,
except at points where it meets the center of symmetry.$)$
It is easy to verify that $S$ is, indeed, a $3$-dimensional totally
umbilic lightlike submanifold. The simplest examples of totally umbilic
lightlike submanifolds constructed in this way are the light cones
in Minkowski space.
\end{example}

%%%%%%%%%%%%%%%%%%%%%%%%%%%%%%%%%%%%%%%%%%%%%%%%%%%%%%%%%%%%%%%%%%%%%%%%%%%%


\begin{thebibliography}{99}
\bibitem{o} B. O'Neill, Semi-Riemannian Geometry, 
  Academic Press, New York, 1983. 
\bibitem{k} D. N. Kupeli, Singular semi-Riemannian geometry,
  Kluwer, Dordrecht, 1996.
\bibitem{db} K. L. Duggal and A. Bejancu, Lightlike submanifolds of
  semi-Riemannian manifolds and applications, Kluwer, Dordrecht,  1996.  
\bibitem{cve} C. M. Claudel, K. S. Virbhadra, and G. F. R. Ellis,
  The geometry of photon surfaces, J. Math. Phys. 42 (2001) 818--838.
\bibitem{fhp} T. Foertsch, W. Hasse, V. Perlick, Inertial forces and 
   photon surfaces in arbitrary spacetimes, Class. Quantum Grav.
   20 (2003) 4635--4652.
\bibitem{v} J. A. G. Vickers, Generalized cosmic strings, 
   Class. Quantum Grav. 4 (1987) 1--9.
\bibitem{a} M. A. Abramowicz, Centrifugal force -- a few surprises,
   Mon. Not. R. Astr. Soc. 245 (1990) 733--746.
\bibitem{pr} R. Ponge and H. Reckziegel, Twisted products in pseudo-Riemannian
   geometry, Geometriae Dedicata, 48 (1993) 15--25.
\bibitem{wa} G. Walschap, Umbilic foliations and curvature, 
   Ill. J. Math. 41 (1997) 122--128.
\bibitem{akk} S. S. Ahn, D. S. Kim D-S, and Y. H. Kim, Totally umbilic
   Lorentzian sumbanifolds, J. Korean Math. Soc. 33 (1996) 507--512.
\bibitem{ho} S. K. Hong, Totally umbilic Lorentzian surfaces embedded
   in $L^n$, Bull. Korean Math. Soc. 34 (1997) 9--17.
\bibitem{sch} A. Schild, Classical null strings, Phys. Rev. D 16 (1977) 
  1722--1726.
\end{thebibliography}
\end{document}